\documentclass[superscriptaddress,aps,showpacs,nofootinbib,twocolumn]{revtex4}
\usepackage{graphicx}
\usepackage{amsmath,amssymb}

\def\MPL{{\it Mod. Phys. Lett.} }
\def\MNRAS{{\it Mon. Not. R. Ast. Soc.} }

\def\PL{{\it Phys. Lett.} }
\def\PR{{\it Phys. Rev.} }

\def\RMP{{\it Rev. Mod. Phys.} }

\def\al{\alpha}
\def\be{\beta}

\def\de{\delta}
\def\ep{\epsilon}

\def\ze{\zeta}
\def\et{\eta}

\def\rh{\rho}

\def\ph{\phi}

\def\ps{\psi}

\def\vev#1{\langle {#1}\rangle}

\def\frac#1#2{{\textstyle{{#1}\over {#2}}}}
\def\goto{\rightarrow}
\newcommand{\lb}{\left (}
\newcommand{\rb}{\right )}

\def\lsim{\mathrel{\rlap{\lower4pt\hbox{\hskip1pt$\sim$}}
    \raise1pt\hbox{$<$}}}
\def\gsim{\mathrel{\rlap{\lower4pt\hbox{\hskip1pt$\sim$}}
    \raise1pt\hbox{$>$}}}
\def\sqr#1#2{{\vcenter{\vbox{\hrule height.#2pt
         \hbox{\vrule width.#2pt height#1pt \kern#1pt
         \vrule width.#2pt}
         \hrule height.#2pt}}}}

\def\beq{\begin{equation}}
\def\eeq{\end{equation}}
\def\beqa{\begin{eqnarray}} 
\def\eeqa{\end{eqnarray}}

\def\laq{\raise 0.4 ex \hbox{$<$}\kern -0.8 em\lower 0.62 ex\hbox{$\sim$}}
\def\gaq{\raise 0.4 ex \hbox{$>$}\kern -0.7 em\lower 0.62 ex\hbox{$\sim$}}

\begin{document}

\title{Chaplygin Inspired Inflation}

\author{O. Bertolami}
\altaffiliation{Email address: orfeu@cosmos.ist.utl.pt}

\affiliation{ Departamento de F\'\i sica, Instituto Superior T\'ecnico \\
Avenida Rovisco Pais 1, 1049-001 Lisboa, Portugal}

\author{V. Duvvuri}
\altaffiliation{Email address: duvvuri@cftp.ist.utl.pt}
 
\affiliation{Centro de F\'\i sica  Te\'orica e de Part\'\i culas, Instituto Superior T\'ecnico \\
Avenida Rovisco Pais 1, 1049-001 Lisboa, Portugal}

\vskip 0.5cm

\date{\today}
\begin{abstract}
We discuss chaotic inflation in the context of a phenomenological 
modification of gravity inspired by the Chaplygin gas
equation of state. We find that all observational constraints can be satisfied provided that the Chaplygin 
scale and the inflaton mass are smaller than $6.9\times 10^{-3}~M$ and $4.7\times 10^{-6}~M$, respectively, 
where $M^2\equiv(8\pi G)^{-1}$ is the reduced Planck mass.
\end{abstract}

\pacs{98.80.Cq, 98.65.Es \hspace{2cm}Preprint DF/IST-1.2006} 

\maketitle
\section{Introduction}

That the universe is undergoing a period of accelerated expansion has become the inescapable fact of cosmology. In order to accommodate an accelerating universe within general relativity, we must postulate the existence of a smooth and unclustered form of stress-energy with a sufficiently negative pressure. Such a component of matter is often referred to as dark energy. The traditional realizations of dark energy are the cosmological constant and quintessence. Whilst being quite satisfactory from the phenomenological viewpoint, neither of these approaches is without serious drawbacks from the viewpoint of fundamental theory. For instance, while we may choose the value of the cosmological constant to fit the data, the most optimistic theoretical estimates of this value remain about $55$ orders of magnitude too large.

Given the challenge of attacking this problem head on, it is not surprising that the search for alternative dark energy candidates is an ongoing one. The Generalized Chaplygin Gas (GCG) \cite{Kamenshchik,Bento1} model is an interesting entry in the current list of candidates. A perfect fluid with a novel equation of state, the GCG mimics the behavior of matter at early-times and that of a cosmological constant at late-times. Thus, in addition to producing the observed late-time acceleration of the universe, the GCG approach allows a unified treatment of dark matter and dark energy. Furthermore, the model has fared quite well when confronted with various phenomenological tests: high precision Cosmic Microwave Background Radiation (CMBR) data \cite{Bento3,Bento3a}, supernova data \cite{Supern,Bertolami1,Bento4}, gravitational lensing \cite{Silva}, gamma-ray bursts \cite{Bertolami2} and cosmic topology \cite{Bento5}. More recently, it has been shown using the latest supernova data that the GCG model is degenerate with a dark energy model with a phantom-like equation of state \cite{Bertolami1,Bento4}. Furthermore, it can be shown that this does not require invoking the unphysical condition of violating the dominant energy condition and does not lead to the big rip singularity in future \cite{Bertolami1}. Structure formation has been studied in Refs. \cite{Bilic,Bento1,Bento6}. A short summary of the results of the various phenomenological tests on the GCG model can be found in Refs. \cite{Bertolami3}. 

As is well known, evidence from the CMBR indicates the early universe underwent an accelerating phase too, viz., the inflationary epoch. Given the attractiveness of the GCG as a dark energy candidate, a natural question to ask is: Can inflation be accommodated within the GCG scenario? This is the question we wish to address in the present work. 

However, we should emphasize that the inflationary model described below is not presented as a more desirable alternative to the conventional ones. Rather, we merely aim to establish the assumptions and extrapolations required to obtain successful inflation in a Chaplygin inspired model.

As mentioned above, the GCG background is described by an exotic equation of state:
\beq
p_{GCG} = - {A \over \rho_{GCG}^{\alpha}} ~~,
\label{eqstate}
\eeq
where $A$ and $\alpha$ are positive constants. The case $\alpha=1$
corresponds to the Chaplygin gas. In most phenomenological studies one considers the range $0 < \alpha \le 1$. Inserting this equation of state into the stress-energy conservation constraint in a Robertson-Walker spacetime leads to the following expression for the energy density \cite{Bento1}:
\beq
\rho_{GCG}=  \left[A + {B \over a(t)^{3 (1 + \alpha)}}\right]^{1 \over 1 +
\alpha}~~.
\label{rhoch}
\eeq
Here, $a(t)$ is the scale-factor of the universe and $B$ is a positive
integration constant. This result demonstrates that, as the universe evolves, the GCG energy density interpolates between that of non-relativistic matter and that of a cosmological constant. This striking property allows the  interpretation of the GCG as dark energy with an admixture of dark matter.

In this paper, we make a key departure from the scenario described above. We will not view Eq. (\ref{rhoch}) as a consequence of the equation of state (\ref{eqstate}), but rather, as arising due to a modification of gravity itself. In particular, we assume that the gravitational dynamics at play during inflation give rise to a modified Friedmann equation of form
\beq
\label{modFried}
H^2={1\over 3M^2}\left[{A+\rh_{\phi}^{(1+\al)}}\right]^{1 \over 1 + \alpha},
\eeq 
where $\rho_{\phi}$ is the energy density of the inflation field (or fields). 
Such a modification is Chaplygin inspired in the sense that it follows from an extrapolation of Eq. (\ref{rhoch}):
\beq
\label{extrap}
\rh_{GCG}=\sqrt{A+\rh_{m}^2}\goto \sqrt{A+\rh_{\phi}^2},
\eeq
where $\rh_{m}$ corresponds to the matter energy density (we set $\al=1$ for simplicity).

That the GCG model may be viewed as a modification of gravity, like so, was first pointed out in Ref. \cite{Barreiro}. It can been shown that the GCG equation of state follows from the generalized Born-Infeld action \cite{Bento1}, and in the case $\alpha=1$, from a single brane setup. In those scenarios, Eq. (\ref{rhoch}) is a consequence of stress-energy conservation for a scalar field on the brane. It is suggestive, then, to view the contribution of the Chaplygin gas to the stress-energy tensor as a brane induced modification to gravity. An analogous situation occurs in the Randall-Sundrum scenario, wherein the standard Friedmann equation is corrected by a $\rho^2$ term \cite{Shiromizu,Binetruy}. 

As it stands, Eq. (\ref{modFried}) constitutes a non-covariant modification of gravity. However, we assume that the effect giving rise to Eq. (\ref{modFried}) preserves diffeomorphism invariance in $(3 + 1)-$dimensions, whence stress-energy conservation follows. Since only the coupling of the inflaton to gravity is modified, the left-hand side of Einstein's equation remains unaltered. Consequently, perturbations of different spins decompose in the conventional manner. However, if the inflaton field equation is to remain unaltered, then the modification we envisage cannot have a Lagrangian description in $(3 + 1)-$dimensions. We must speculate that the effect has a higher-dimensional origin. While this is indeed a drawback, our goal here is purely phenomenological and a derivation of the model from a fundamental theory is beyond the scope of this paper.

While modifications of gravity have been proposed in the context of braneworld scenarios (see e.g. \cite{Rubakov,Langlois1,Maartens1} for reviews), non-covariant modification in particular have been considered in \cite{Freese, Arkani-Hamed, Dvali, Mersini}. 
These gravity modifications can imply in important changes in the early universe dynamics and affect, for instance, inflation \cite{Maartens2,Bentoinf1,Bentoinf2,Bentoinf3}, or introduce late time changes in the expansion rate of universe, and in some instances, successfully account for the late time accelerated expansion without the need to introduce dark energy. It is quite suggestive to think that these two crucial occurrences in the history of the universe have a similar underlying dynamics. Having this purpose in mind, some models have been put forward where inflation and the late time accelerated expansion are due to the same mechanism \cite{Peebles,Dimopoulos,Rosenfeld}.

For simplicity, we restrict attention to the Chaplygin gas inspired 
model (the $\alpha=1$ case) in this paper. Hence, 
the modified Friedmann equation reads
\beq
\label{modFriedmann}
H^2={1\over 3M^2}\sqrt{A+\rh_{\phi}^2}.
\eeq 
As an aside, we remark that when applied to late universe cosmology, the $\alpha=1$ case is consistent with most 
of the phenomenological constraints, but requires a rather low Hubble constant, $h \lsim 0.64$ \cite{Bento3a,Bertolami3}, in order to fit the CMBR data.

To be specific, we shall consider chaotic inflation \cite{Linde1} in the context of our modified Friedmann equation. In other words, the energy density in Eq. (\ref{modFriedmann}) will be one associated with an inflaton, $\phi$, which is chosen to 
have a canonical kinetic term and a potential that is dominated by a mass term $V=m^2\ph^2/2$. Then, the Klein-Gordon equation for the inflation is given by
\beq
\label{scalarEOM}
\ddot\ph+3H\dot\ph+V'(\ph)=0.
\eeq

\section{The Model}

An inflationary phase is one wherein the universe undergoes an accelerating expansion; 
i.e., the scale factor satisfies $\ddot a > 0$. Inflation ends when this condition is violated. 
In general relativity, this inequality is achieved when the pressure and energy density of the 
universe satisfy 
\beq
\label{GR_acc}
p < -\rh/3~.
\eeq

In particular, when the stress-energy of the universe is 
scalar field dominated, the onset of inflation is brought about by $\dot\ph^2  \ll V$, 
whence $p=-\rh$. The analogous criterion in Chaplygin inflation can be obtained by 
differentiating equation (\ref{modFriedmann}) and using $\dot\rh_{\phi}=-3H(\rh_{\phi}+p_{\phi})$. It is
\beq
\label{acc_ineq}
\rh_{\phi}(\rh_{\phi}+3p_{\phi})<2A~.
\eeq
While the inflaton satisfies this inequality, we have $\ddot a > 0$. We should note that equation (\ref{GR_acc}) 
is recovered in the limit $A\goto 0$. 
This, of course, is the limit wherein we pass to general relativity, too.

As we have remarked above, the inflaton in our model obeys conventional scalar field 
dynamics. Thus, in the slow-roll approximation ($\dot\ph^2  \ll V$ and $\ddot\phi \ll V'$) 
equations (\ref{modFriedmann}) and (\ref{scalarEOM}) reduce to 
\beqa
\label{slowrollEOM}
H^2&\simeq& {1\over 3M^2}\sqrt{A+V^2},\\
\dot\ph&\simeq&-{\sqrt{3}M V'\over 3(A+V^2)^{1/4}},
\eeqa
where the symbol $\simeq$ denotes equality during the slow-roll regime.
Self-consistency of the slow-roll approximation requires that the following inequalities be satisfied during inflation:
\beqa
\label{slowroll}
\ep &\equiv& -{\dot H\over H^2}={M^2 V(V')^2\over 2(A+V^2)^{3/2}}\ll 1 ,\\
\de &\equiv& -{\ddot\ph \over H\dot\ph}={M^2 V''\over (A+V^2)^{1/2}}-{\ep}\ll 1 ,\\
\et &\equiv& \ep+\de ={M^2 V''\over (A+V^2)^{1/2}} \ll 1.
\eeqa

In general relativity, $\ep\ll 1$ is tantamount to the slow-roll condition $\dot\ph ^2 \ll V$. 
Evidently, this is not so in the case at hand. Rather, we have 
\beq
\label{pot_domination}
{\dot\phi ^2\over V}={2\over 3}\lb 1+ {A\over V^2}\rb\ep,
\eeq
which is small if and only if $A/V^2 \lesssim \mathcal {O}(1)$. Indeed, 
this condition remains valid throughout the Chaplygin regime. In fact, 
we may view this condition as signaling the onset of the Chaplygin inflation effect.

As a first step toward evaluating various inflationary observables, we require $\ph_e$, the value 
of the inflaton field amplitude when inflation ends. This can be obtained by solving (\ref{acc_ineq}) for $\ph$, 
when $\ddot a > 0$ ceases to be true. In the slow-roll regime, we can write (\ref{acc_ineq}), using Eq.(10), as 
\beq
\label{acc_ineq2}
V^2(V')^4< {4\over M^4}(A+V^2)^3.
\eeq
Specializing to the case $V=m^2\ph^2/2$ obtains
\beqa
\label{acc_ineq3}
m^{12}\ph_e^{12}+\lb 12Am^8-4M^4 m^{12}\rb\ph_e^8 \nonumber\\
+ 48A^2 m^4 \phi_e^4+ 64 A^3= 0. \qquad
\eeqa
Solving this equation for $\ph_e$ yields
\beq
\label{phi_e}
\phi_e= 1.4~M,
\eeq
where, while solving the cubic equation for $\ph_e^4$, we have retained only terms involving the highest powers of $m M/A^{1/4}$. 

The number of e-folds during inflation is given by
\beq
\label{efolds}
N(\ph_b\goto\ph_e)\simeq -\int_{\ph_*}^{\ph_e}d\ph{\sqrt{A+V^2} \over M^2V'},
\eeq
where $\ph_*$ is the field amplitude at the time of horizon-crossing for the scale in question. 
For a quadratic potential, evaluating the integral obtains
\beqa
\label{chaoticefolds}
N(\ph_b\goto\ph_e)={1\over 2M^2}\sqrt{{\ph_b^4\over 4}+{A\over m^4}}-{1\over 2M^2}\sqrt{{\ph_e^4\over 4}+{A\over m^4}}
\nonumber\\-{1\over 2M^2}\sqrt{{A\over m^4}}
\log{{\ph_e^2\lb\sqrt{A\over m^4}+\sqrt{{\ph_b^4\over 4}+{A\over m^4}}\rb}\over\ph_b^2\lb\sqrt{A\over m^4}
+\sqrt{{\ph_e^4\over 4}+{A\over m^4}}\rb}.
\eeqa
Since $\ph_e \sim M$, and $\ph_* > \ph_e$, the first term on the right-hand side of the above expression is the dominant one. In order to ascertain if a sufficient amount of inflation can occur, we substitute $\ph_e=1.4~M$ into the above expression and solve solve for $\ph_*$ in terms of $N$. For large $N$, upon dropping the logarithm, we obtain 
\beq
\label{phi_*}
\phi_*=2\sqrt{N}M.
\eeq

\section{Observational Bounds}

In this section, we will ascertain how Chaplygin inflation fares when confronted with 
CMBR data. In particular, we will obtain the power spectra of scalar and tensor perturbations 
to the metric in Chaplygin inflation. Toward this end, we compute the gauge invariant quantity
\beq
\ze=\ps+H{\de\rh \over \dot\rh},
\eeq
where $\ps$ is the gravitational potential. On slices of uniform density $\ze$ 
reduces to the curvature perturbation. A key attribute of $\ze$ is that it is nearly constant on 
super-horizon scales. This fact, being a consequence of stress-energy conservation, does not depend 
on the gravitational dynamics. Thus it remains unaltered in Chaplygin inflation. 

The amplitude of scalar perturbations can be obtained from the two-point function of $\ze$:
\beq
A_S^2={4 \over 25}\vev{\ze^2},
\eeq
where we have adopted the normalization convention of \cite{Lidsey}. 
It can be shown that on super-horizon scales, the curvature perturbation on 
slices of uniform density is equal to the comoving curvature perturbation. 
Thus, in a spatially flat gauge, we have
\beq
\ze=H {\de\ph \over \dot\ph},
\eeq
Since $\de\phi$ is ``frozen in'' on super-horizon scales, the expression 
above can be evaluated around the time of horizon crossing, whence the inflation 
fluctuation is given by the Gibbons-Hawking temperature of the de Sitter space event 
horizon. Using $|\de\ph|=H/2\pi$, we arrive at 
\beq
\label{scalar_amp}
A_S^2={1\over 75\pi^2M^6}{(A+V^2)^{3/2}\over(V')^2}|_{k=aH}.
\eeq
As a check, we may note that in the limit $A\goto 0$ the familiar expression 
$A_S^2 \propto H^2/\ep|_{k=aH}$ is recovered.

Now specializing to a quadratic potential and introducing the observational bound from COBE, $A_S=2\times 10^{-5}$, we have
\beq
\label{scalar_amp2}
{1\over 75\pi^2 \be^2 M^2 m^4}\lb{A\over M^4}+{\be^4 m^4\over 4}\rb ^{3/2} < 4 \times 10^{-10},
\eeq
where $\be = 2\sqrt{N}M$. Defining $x=m^4/M^4$ and $y=A/M^8$, we can rewrite this bound as
\beq
x^{-1}\lb y + {\be ^4\over4}x \rb ^ {3/2} < 3\be ^2\times 10^{-7}~.
\eeq
Taking $\be=14.8$, which corresponds to the large angle anisotropies measured by COBE, we arrive at the following 
bounds on the Chaplygin inflation scale and the mass of the inflaton:
\beqa
A^{1/8}&<& 6.9 \times 10^{-3}M,\\
m&<&4.7 \times 10^{-6}M.
\eeqa
This is the central result of this paper; it demonstrates the viability of 
our model vis-a-vis observational bounds from CMBR anisotropies. Furthermore, it is 
pleasing to note that while the inflaton mass is close to that required by conventional 
inflationary models, the hierarchy between the Planck and Chaplygin scales is not very large.

For completeness, let us give the spectral indices, which quantify the scale dependence of the power spectra. When $A$ and $m$ saturate the bounds given above, the scalar spectral index is given by $n_S=1+[2\et-6\ep]|_{k=aH}=0.992$.  Lowering the value of the Chaplygin scale somewhat, say by taking $A^{1/8}=4\times 10^{-3}$, obtains $n_S=0.965$. This is in accord with the latest Wilkinson 
Microwave Anisotropy Probe (WMAP) data \cite{WMAP}, too. The tensor spectral index is given by $n_T=-2\ep|_{k=aH}=-1.8\times 10^{-2}$.


The tensor spectrum amplitude is also readily obtained. To be consistent with the convention adopted in 
defining $A_S$ above, we define it as \cite{Lidsey}
\beq
A_T^2={H^2\over 50\pi^2M} 
={\sqrt{A+V^2}\over150\pi^2M}.
\eeq
It follows that the tensor-to-scalar amplitude ratio is \cite{Lidsey}
\beq
\label{ratio}
\left({A_T \over A_S}\right)^2
=-{(A+V^2)^{1/2}\over V}{n_T\over 2},
\eeq
where once again we may note that the familiar expression 
$A_T^2/A_S^2=\ep$ is recovered in the limit $A\goto 0$. In our model, the right-hand side of Eq. (\ref{ratio}) takes on the 
value $9.1\times10^{-3}$.

\section{Discussion and Conclusions}

In this work we have considered chaotic inflation in the context 
a phenomenological modification of gravity 
inspired by the Chaplygin gas equation of state. We find 
that observational bounds on CMBR anisotropies are satisfied provided that the  
Chaplygin scale and the inflaton mass are smaller than $6.9\times 10^{-3}~M$ and
$4.7\times 10^{-6}~M$, respectively. Also, we find that scalar fluctuations are characterized by a spectral index of $0.965$, while the spectral index of gravitational waves is $-1.8\times 10^{-2}$. The ratio of the amplitude of gravitational waves to that of fluctuations in the energy density is $9.1\times 10^{-3}$. These results  are consistent with available WMAP data \cite{WMAP}.

We have not yet addressed reheating and the transition to standard cosmology in our model. A proper discussion of these issues requires a complete knowledge of the underlying gravitational dynamics and is beyond the scope of this paper. However, we give a plausibility argument by analogy with braneworld cosmology. Due to brane effects, the dynamics is ruled by Eq. (\ref{modFriedmann}), even 
though a term linear, $\rh_{\phi}$, is present in the Friedmann equation. 
During inflation, the term quadratic in $\rh_{\phi}$ clearly 
dominates. As the inflaton rolls to the minimum of the potential, the linear term comes into play. As stated in the introduction, we assume that the dynamics responsible for Eq. (\ref{rhoch}) ceases to operate when inflation ends. 
Thereafter, we are left with a conventional Friedmann equation and a large cosmological constant, $\sqrt{A}$; cancellation of the later 
presents the conventional cosmological constant problem.

Subsequently, the inflaton reaches the bottom of its potential and reheating follows as in conventional chaotic inflation. Assuming a fully efficient thermal conversion , and that the number of degrees of freedom is about 150, one can estimate the reheating temperature in our model to be $T_{RH}= 1.2 \times 10^{-3}~M$.

For sure, inflation in the context of the our model is compatible with the 
dynamics of some other field which may 
give rise to another inflationary period, likewise in the 
so-called hybrid inflationary models \cite{Linde2}.

\vskip 0.2cm
\begin{acknowledgments}

\vskip 0.2cm

\noindent 
The work of O.B. is partially supported by Funda\c c\~ao para a
Ci\^encia e a Tecnologia (FCT, Portugal) under the grant POCI/FIS/56093/2004.
The work of V.D. is fully supported by the same grant.

\vskip 0.2cm
\end{acknowledgments}


\end{document}